# LymphAtlas : A Unified Multimodal Lymphoma Imaging Repository Delivering AI-Enhanced Diagnostic Insights

Jiajun Ding[a], Beiyao Zhu[b], Xiaosheng Liu[c], Lishen Zhang[a], Zhao Liu[a*]

[a] *School of Design, Shanghai Jiao Tong University, Shanghai, PR China*

[b] *Shanghai Ninth People's Hospital, Shanghai Jiao Tong University School of Medicine, Shanghai, P.R. China.*

[c] *Department of Nuclear Medicine, Fudan University Shanghai Cancer Center, Shanghai 200032, China*

## Abstract

This study integrates PET metabolic information with CT anatomical structures to establish a 3D multimodal segmentation dataset for lymphoma based on whole-body FDG PET/CT examinations, which bridges the gap of the lack of standardised multimodal segmentation datasets in the field of haematological malignancies. We retrospectively collected 483 examination datasets acquired between March 2011 and May 2024, involving 220 patients ; all data underwent ethical review and were rigorously de-identified. Complete 3D structural information was preserved during data acquisition, preprocessing and annotation, and a high-quality dataset was constructed based on the nnUNet format. Through systematic validation of the preprocessing, annotation quality and segmentation algorithm, models trained on this dataset demonstrate highly accurate, robust and reproducible lymphoma lesion segmentation in PET/CT—confirming its applicability and stability for precise segmentation and quantitative analysis. By deeply fusing metabolic and anatomical information, the dataset markedly enhances characterization of tumour morphology, location and metabolic activity, providing a strong foundation for early diagnosis, clinical staging, personalized treatment and the advancement of deep learning–based automated segmentation and precision medicine.

## 1. Introduction

Lymphoma is a class of clonal proliferative diseases originating from B cells, T cells, or natural killer cells, and its mechanism of development is closely related to genetic susceptibility, viral infections (e.g., EBV, HIV), and immune dysregulation [1][2]. In recent years, the incidence of lymphoma has shown a rising trend due to the increase in environmental exposures and associated pathogen infections, making it an important public health problem worldwide.

Globally, the annual number of new cases of lymphoma has climbed from about 566,000 in 2012 to about 602,000 in 2022, with non-Hodgkin's lymphoma (NHL) accounting for 553,000 new cases and 250,475 deaths in 2022, which is a significantly higher growth rate than that of most solid tumors[3][4]. Lymphoma is mainly divided into Hodgkin lymphoma (HL) and non-Hodgkin lymphoma (NHL), of which NHL accounts for about 92% of all new cases of lymphoma, and the age-standardized mortality rate of NHL in 2022 reached 2.6/100,000, with a higher mortality rate; although the prognosis of HL is better, the new-onset patients are mainly concentrated in young and middle-aged people aged 15-35 years, and the disease has a socioeconomic impact of the disease is significant [5]. The clinical presentation of lymphoma is highly heterogeneous, ranging from asymptomatic to extensive multisystem invasion. The most common and typical first manifestation is painless lymph node enlargement, often involving the neck, axilla or groin, with progressive enlargement of lymph nodes, firm texture, and no tenderness on palpation. In addition to lymph nodes, lymphoma may occur with extensive organ infiltration, including splenomegaly, hepatomegaly, gastrointestinal lesions, skin nodules and central nervous system involvement, which may be clinically manifested as dysfunction or localized symptoms in the corresponding organs[6][7]. Because the biological behavior and prognosis of lymphoma vary greatly with subtype and early symptoms are often atypical or easily overlooked, the need for early diagnosis, precise typing and individualized treatment is particularly urgent.

In the imaging examination of lymphoma, traditional single-modality imaging methods mainly include computed tomography (CT) and positron emission tomography (PET).CT imaging has obvious advantages in anatomical structure display and lesion localisation due to its high spatial resolution, but it is deficient in reflecting the biological activity of the tumour; whereas PET imaging can provide metabolic information of tumours by detecting the distribution of tracers such as 18F-FDG that can provide metabolic information of tumours, but due to the lower spatial resolution, it is often difficult to accurately determine the anatomical location of the lesion, thus leading to the obvious limitations of single modality in clinical diagnosis [8]. In current academic circles, although single-modality imaging technology has achieved more mature applications in the diagnosis of lymphoma, multimodal image segmentation research for this disease still faces many difficulties. Firstly, as lymphoma exhibits large heterogeneity in different imaging modalities, its lesion morphology, location and metabolic features are difficult to be comprehensively portrayed by a single modality, leading to high uncertainty and challenge in image segmentation tasks [9][10].

To overcome the shortcomings of unimodal imaging, PET/CT multimodal imaging technology has emerged, which combines the functional information of PET with the high-resolution anatomical images of CT to achieve precise localisation and quantitative analysis of lesions, significantly improving the accuracy of disease staging and efficacy assessment [[11][12]. However, existing multimodal imaging datasets are mainly focused on other fields such as brain tumours, e.g., the BRATS dataset has been widely used in brain tumour segmentation, but no publicly available and standardised multimodal segmentation dataset has been made available for research in lymphoma.

This current deficiency seriously restricts the development of automated segmentation algorithms and accurate diagnostic models based on multimodal imaging technology, which in turn affects the clinical staging, efficacy assessment and implementation of personalised treatment for lymphoma. Based on the above background, the study of constructing a high-quality multimodal segmentation dataset integrating PET and CT data is of great theoretical and clinical significance. Combining CT and PET multimodal imaging technologies for accurate segmentation and quantitative analysis of tumours not only improves diagnostic accuracy, but also provides powerful support for radiotherapy planning, efficacy assessment and targeted therapy. This will not only help to improve the accuracy and robustness of image segmentation, but also provide valuable data support for promoting deep learning applications and precision medicine development in this field.

In summary, to cope with the lack of existing multimodal imaging datasets in lymphoma research, this study follows the trend, aiming to fill this data gap and promote the development of automated segmentation algorithms and accurate diagnostic models. We constructed a high-quality multimodal segmentation dataset integrating PET and CT data through retrospective collection and rigorous preprocessing process, which provides solid data support for accurate segmentation and quantitative analysis of lymphoma. Based on this, the main contributions of this study are as follows:

- **Constructing high-quality multimodal segmentation datasets:**

In this study, we retrospectively collected whole-body FDG-PET/CT examination data obtained in routine clinical practice between March 2011 and May 2024, and the dataset covered 220 lymphoma patients and 483 examinations, especially including imaging data of 106 non-Hodgkin's lymphoma and 42 Hodgkin's lymphoma patients. Utilizing real clinical data, this study completed a series of data acquisition, preprocessing, and annotation, and constructed a multimodal segmentation dataset according to the requirements of the nnUNet [13] format.

- **Rigorous technical validation and analysis of results:**

By systematically evaluating the data preprocessing process, the annotation quality and the results of the segmentation algorithm, this study rigorously validates the applicability of the dataset in automated segmentation tasks and ensures the validity and reliability of the constructed dataset in precision diagnosis and clinical decision support.

- **Advancing Deep Learning and Precision Medicine:**

By leveraging high-quality multimodal datasets, this study achieves effective fusion of PET and CT image information, which significantly improves the accurate portrayal of morphological and metabolic features of lymphoma lesions. This not only provides powerful support for early diagnosis, staging and efficacy assessment of the disease, but also opens up a new way for the development of automated image segmentation and precision medicine strategies based on deep learning, which has important theoretical and clinical application value.

The remainder of this study is organised as follows: section 2 describes the specific work; section 3 contains the instructions for use; and section 4 concludes.

# 2. Methods

## 2.1 Data Collection

The anonymised data collection in this study was approved by the Ethics Committee and the Data Security and Privacy Review Committee of Fudan University Cancer Hospital. We retrospectively collected data on whole-body FDG-PET/CT examinations obtained during routine clinical consultations at this hospital between March 2011 and May 2024, involving 220 patients and 483 examinations. The collected data were derived from actual clinical needs, with high representativeness and authenticity, providing a solid data foundation for the development of subsequent diagnostic models and efficacy evaluation.

Among these 483 examinations, we focused on screening the imaging data of patients with a diagnosis of lymphoma for the purpose of exploring the imaging characteristics of FDG-PET/CT under different pathological types and treatment stages. Specifically, the dataset contained whole-body imaging data of 106 patients with non-Hodgkin's lymphoma and 42 patients with Hodgkin's lymphoma , whose age distribution ranged from 15 to 76 years old, with a balanced male-to-female ratio. The collected data not only included the baseline images of the patients at the time of diagnosis, but also covered the patient number, clinical information, and clinical diagnosis, as detailed in Table 1.

**Table 1**: Table of clinical information of patients

| No. | Date | Gender | Age | ID | Clinical Data | Clinical Diagnosis |
|---|---|---|---|---|---|---|
| 1 | 2022 0318 | Male | 41 | 1136 5954 | Repeatedly coughing for more than 5 times, on December 9, 2021, an external CT scan showed a space occupying lesion in the upper and middle lobe of the right lung, suggesting malignancy possibly accompanied by enlarged mediastinal and hilar lymph nodes. On December 9, 2021, an external CT scan showed a space occupying lesion in the upper and middle lobe of the right lung, suggesting malignancy possibly accompanied by enlarged mediastinal and hilar lymph nodes. was performed in our hospital, and the pathology showed (right neck) lymph node classic Hodgkin's lymphoma with nodular sclerosis. | Nodular sclerosis Hodgkin's disease |

## 2.2 Data Acquisition

Currently, all F-FDG PET/CT examinations were performed at the Cancer Hospital of Fudan University according to a standardised acquisition protocol using the Biograph mCT Flow

PET/CT scanner from Siemens Healthineers, Germany, and in strict compliance with the International Guidelines for Oncology Imaging FDG-PET/CT (EANM version 2.0). Patients were fasted for ≥6 hours and fasting blood glucose was confirmed to be <11 mmol/L before injection of ^18F-FDG. 3.7 MBq/kg of body weight was administered intravenously, and patients rested quietly for approximately 60 minutes after injection before image acquisition.

The scanning range was extended from the mid-thigh to the top of the skull, and the first low-dose CT acquisition was performed (120 kV, 140 mA, 3 mm layer thickness, no contrast, and the patient breathed calmly) for attenuation correction and anatomical alignment, followed by a continuous bed motion PET scan (speed 2.0 mm/s), which provided full coverage of the whole body's functional and metabolic information.PET images were reconstructed using iterative reconstruction combined with Gaussian filtering (the specific number of iterations and filter width were adjusted according to the clinical needs). The PET images were reconstructed using iterative reconstruction combined with Gaussian filtering (the exact number of iterations and filter width were adjusted according to clinical needs), and a homogeneous matrix of voxels of approximately $2.04 \times 2.04 \times 3$ $mm^3$ was generated, which provided a high-quality data base for quantitative analysis and multi-temporal comparisons.

## 2.3 Data Processing

To ensure the consistency of FDG-PET and CT bimodal images in spatial coordinates and subsequent quantitative analyses, a rigorous and systematic preprocessing process was implemented in this study after data acquisition. The process mainly includes data reading and metadata extraction, preliminary spatial normalization, resampling, origin and orientation correction, image alignment, and subsequent processing in the following steps:

### 2.3.1 Data Loading and Metadata Extraction

Firstly, the original DICOM data of each case were read by a specially written programme to obtain PET and CT image information respectively. The extracted metadata include, but are not limited to, Image Size, Voxel Size/Spacing, Origin, Direction Cosines, Field of View (FOV), and Patient Position. This step provides the necessary parameters for subsequent image unification and fusion in physical space.

### 2.3.2 Preliminary Spatial Normalization

Based on the extracted DICOM metadata, PET and CT data are mapped into their respective physical space coordinate systems. The specific method is to construct the spatial coordinate mapping of each image using the voxel size, origin and direction matrices, through which the positional localisation of each pixel point within the image in the real physical space is achieved, thus laying a solid foundation for the subsequent cross-modal data alignment.

### 2.3.3 Resampling

Due to the different scanning parameters used in the acquisition process of PET and CT

images, there is a significant difference between their image size and voxel size. In order to eliminate this difference, in this study, CT images were used as the reference modality, and PET data were resampled by interpolation methods (e.g., cubic interpolation) to obtain the same voxel dimensions and image matrix dimensions as CT. During the resampling process, the origin information of the PET data is simultaneously adjusted to ensure the initial alignment of the two modalities in terms of spatial location.

### 2.3.4 Origin and orientation correction

While resampling is performed, the PET and CT images are corrected based on the origin and orientation information recorded in the DICOM metadata to ensure that the coordinate starting points and orientations of both are consistent in physical space. By adjusting the orientation matrix and correcting the origin, the base consistency of different modal images under rigid spatial transformation is ensured at this stage, which provides support for subsequent image fusion and quantitative analysis.

### 2.3.5 Image Alignment

Although the PET and CT images already have a high spatial consistency after the aforementioned resampling and correction, a small amount of alignment bias may still be introduced due to factors such as patient motion during the acquisition process. In order to further improve the image alignment accuracy, this study adopts a rigid alignment algorithm, which is combined with a non-rigid alignment technique to further fine-tune the calibrated data when necessary. Mutual Information is used as an optimisation metric during the alignment process to achieve optimal alignment between cross-modal images and to ensure that the lesion area corresponds accurately between the two modalities.

### 2.3.6 Post-processing

A further post-processing step, including intensity normalization, noise filtering and automatic/semi-automatic extraction of regions of interest (ROIs), is performed on the dataset after image alignment. This step is designed to enhance data quality and stability for subsequent applications such as quantitative analysis, computer-aided diagnosis and clinical treatment evaluation. All pre-processing steps are subject to stringent quality control to ensure that each process point meets predetermined standards.

Through the above pre-processing procedure, we have achieved the consistency of PET and CT bimodal images in terms of size, resolution, origin and direction, thus providing a solid foundation for the subsequent image fusion, lesion segmentation and quantitative analysis, and effectively ensuring the spatial consistency of the multimodal dataset and the value of its clinical application.

## 2.4 Data labelling

All imaging examinations were performed by two nuclear medicine physicians with extensive clinical experience under a rigorous clinical evaluation process. Based on the preliminary clinical diagnosis and imaging report, the research team identified the lymphoma region as the region of primary interest (ROI) to explore its FDG uptake characteristics and clinical application value. The specific process is as follows: first, after image pre-processing and quality control, two experts used ITK-Snap software to perform detailed layer-by-layer segmentation of the aligned FDG-PET/CT data, respectively. During the annotation process, the experts identified lesion areas where FDG uptake was significantly higher than the blood pool level by comparing the PET with the corresponding CT images (using side-by-side display or superimposed display), and accurately segmented the lesion areas according to the established operation guidelines to generate the corresponding 3D binary masks. If the two experts disagree on the boundary or extent of a lesion during segmentation, the case is submitted to a third senior radiation oncologist for review. This expert will make an independent judgement and confirm the segmentation result based on clinical experience and image features, thus ensuring the highest accuracy and consistency of the entire data annotation process.

The DICOM and mask data after completion of the annotation were strictly anonymised to ensure patient privacy and data security. The whole annotation process and quality control measures provide solid data support for subsequent quantitative analysis, treatment response assessment and related clinical research.

Figure 1 below demonstrates the PET and CT patient images, each of which presents cross-sectional views in three directions (axial, coronal and sagittal) to visualise the three-dimensional structure and spatial distribution of the lesion area. These multi-angle cross-sectional images help clinicians observe image details from multiple viewpoints, providing a powerful reference for lesion localisation, boundary identification and morphological analysis.

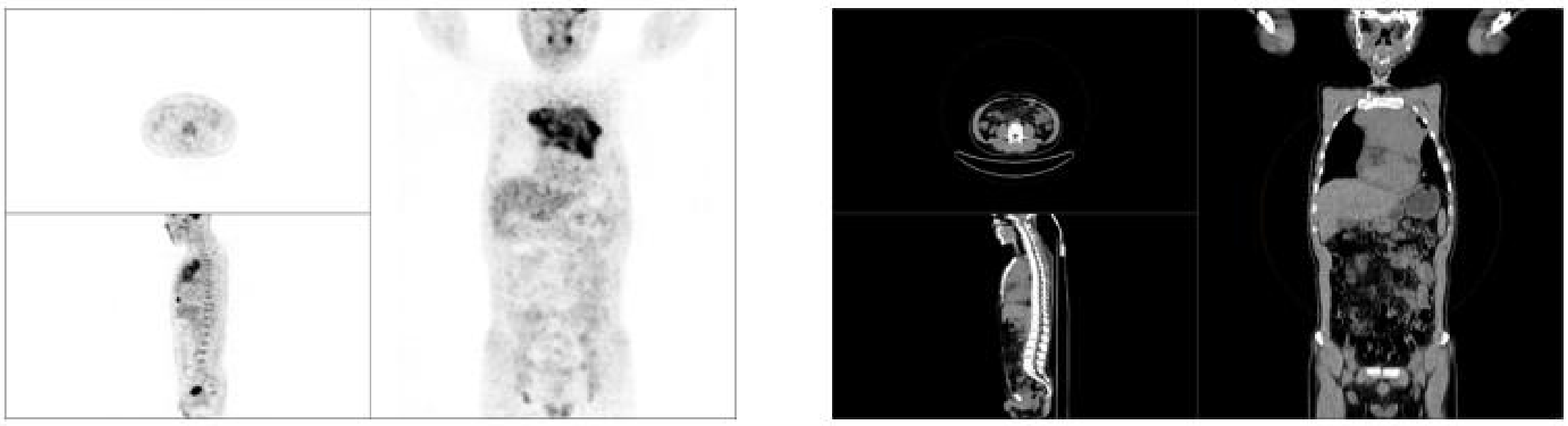

**Figure 1**: Image presentation of the dataset, with the patient's PET image on the left and the patient's CT image on the right; each image contains transverse views in axial, coronal and sagittal planes.

Figure 2 below shows the multimodal images and labelling results of the patient's PET and CT. On the left side of the figure is the PET and CT information fusion map, which visualises the anatomical structure and metabolic activity of each region in the image; in the middle is the CT image superimposed with red labels, which clearly annotates the specific location and morphology of the lesion region; and on the right side is the PET image superimposed with red labels, which highlights the degree of activity of the lesion region in the metabolic distribution.

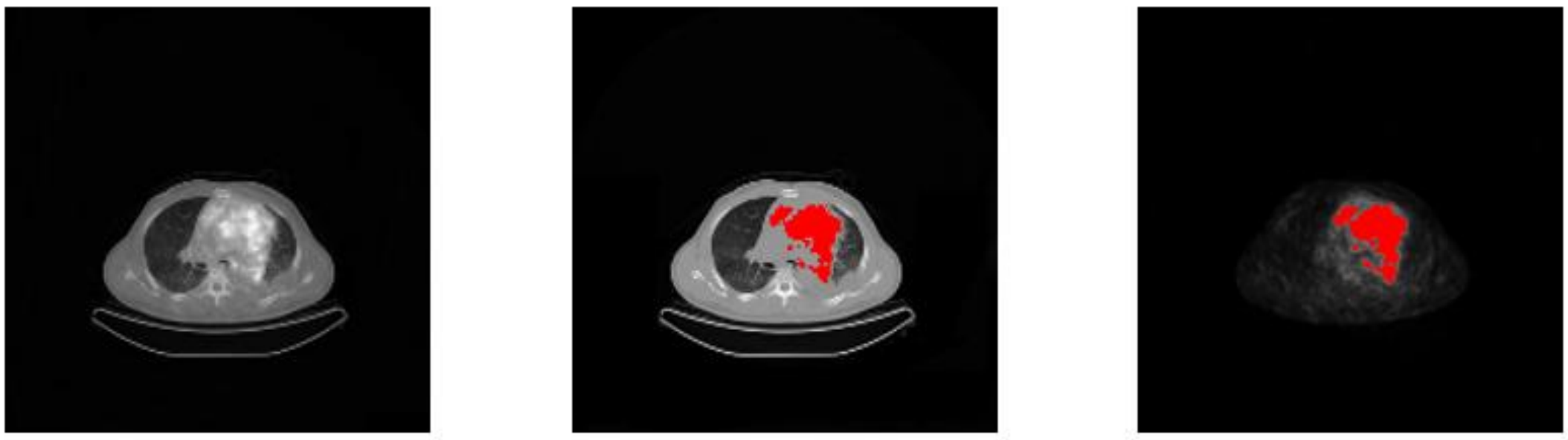

**Figure 2**. Schematic of the dataset annotation display: the left side shows images fusing PET and CT information; the middle shows CT images with red labels superimposed; the right side shows PET images with red labels superimposed.

## 2.5 Description of the dataset.

After a pre-processing procedure and rigorous multi-step calibration, this study converted the bimodal FDG-PET/CT data into the data structure recommended by nnU-Net for subsequent deep learning-based image segmentation and quantitative analysis. The raw data were acquired in DICOM format, and although the DICOM format contains comprehensive metadata information and image details, its file structure is cumbersome and large, which is not conducive to high-speed read/write and batch processing of large-scale data. In order to overcome these shortcomings, we used a professional data conversion tool to convert the data format to Niigz format (.nii.gz), which has the advantages of simple structure, small file size, fast random access speed, etc. At the same time, the conversion process eliminates the differences in encoding methods between different devices and manufacturers, thus ensuring the standardisation and uniformity of the image data of each case after data pre-processing. standardisation and uniformity of the image data of each case after data preprocessing.

After completing the format conversion and data standardisation, we constructed the dataset based on the requirements of the nnU-Net framework. The dataset strictly follows the recommended file organisation of nnU-Net, i.e., it includes the basic parts of training images, training labels and test images. The PET and CT image data and corresponding segmentation labels of each case are stored with uniform resolution, image size, and spatial alignment information to ensure that the different modality data reach an exact match under the physical coordinate system. Specifically, this study covers 42 cases of Hodgkin's lymphoma (HL) and 106 cases of non-Hodgkin's lymphoma (NHL), each of which contains PET/CT dual-modality images of the whole body range except the head and finely segmented lesion areas. The following figure shows the directory format of the constructed dataset, where the function of each file and folder is described below:

- dataset.json: defines the basic information of the dataset, including the dataset name, modal number and name mapping (e.g., "0": "CT", "1": "PET"), label meanings (e.g., "0": "background", "1": "lesion"), and a list of training and testing samples.
- imagesTr/: stores the training set image data. The bimodal data of each case are distinguished by the file suffix, where _0000.nii.gz denotes CT images and _0001.nii.gz denotes PET images; a total of 42 cases of HL and 106 cases of NHL are included in the image data.

- labelsTr/: stores the training set segmentation labels. Each label file corresponds to the number of the corresponding case in imagesTr and is stored as a 3D binary mask for labelling the lesion area.
- imagesTs/: reserved test set image data storage directory, currently there is no label file, the subsequent can be supplemented as needed or used for independent testing.

```
Dataset_Lymphoma
├── dataset.json
├── imagesTr
│   ├── HL_001_0000.nii.gz
│   ├── HL_001_0001.nii.gz
│   ├── ...
│   ├── HL_042_0000.nii.gz
│   ├── HL_042_0001.nii.gz
│   ├── NHL_001_0000.nii.gz
│   ├── NHL_001_0001.nii.gz
│   ├── ...
│   ├── NHL_106_0000.nii.gz
│   ├── NHL_106_0001.nii.gz
├── imagesTs
│   ├── （Test image）
└── labelsTr
    ├── HL_001.nii.gz
    ├── ...
    ├── HL_042.nii.gz
    ├── NHL_001.nii.gz
    ├── ...
    ├── NHL_106.nii.gz
```

**Figure 3**: Structure of the dataset catalogue

The standardised data structure not only improves the adaptability and training efficiency of deep learning models, but also enhances the reproducibility of research results and the convenience of data sharing, providing a solid data foundation for joint multimodal image analysis and clinical applications.

Combining the above processing and data format conversion, the constructed standardised dataset not only maintains the consistency of multimodal images in terms of size, resolution, origin and orientation, but also makes full use of the high efficiency of the Niigz format in data storage and reading, and at the same time meets the stringent requirements of the nnU-Net framework for data organisation, which will lay a solid foundation for the subsequent deep-learning-based lesion segmentation, quantitative analysis, and clinical research. It lays a solid foundation for subsequent deep learning-based lesion segmentation, quantitative analysis and clinical research.

## 2.6 Characteristics of the dataset.

Carrying on the data preprocessing and standardisation process described above, the dataset of this study fully reflects the unique disease characteristics of lymphatic system tumours while meeting the needs of efficient training of deep learning models, with the following main features:

- **3D full body image segmentation**

The dataset covers the whole body except the skull, and is able to capture the characteristics of lymphatic system tumours such as scattered, multiple and irregular morphology throughout the body, reflecting the disease characteristics of complex tumour contours, fuzzy margins and wide distribution.

- **Complementary advantages of PET/CT dual modality**

CT images demonstrate anatomical structures in detail and help identify the boundaries between the tumour and surrounding tissues; PET images provide information on tumour metabolism and reveal lesion activity. Both images are finely aligned to achieve strict agreement in physical coordinates, providing multi-dimensional support for fine segmentation and biomedical interpretation of complex tumour patterns.

- **Strict data standardisation and format conversion**

Converting the original DICOM data to NIfTI format and organising the data according to the directory structure recommended by nnU-Net not only ensures the consistency of different modal images in terms of resolution, size and alignment, but also lays a solid data foundation for subsequent multimodal lesion segmentation and quantitative analysis, while facilitating the data sharing and reproduction of results.

In summary, this dataset fully integrates the advantages of 3D whole-body imaging, the specificity of lymphatic system tumour distribution and morphology, and the complementary information of PET/CT dual-modality, which not only possesses clear biomedical interpretability, but also meets the requirements of deep learning for data uniformity and efficient processing, providing reliable data support for the research of fine segmentation and quantitative analysis of complex lesions.

# 3. Usage Notes:

## 3.1 Instructions for use

This data is available at https://github.com/SuperD0122/LymphAtlas- . Meanwhile, this dataset supports a variety of medical imaging software for visualisation and analysis, and users are recommended to use open-source medical image viewers (e.g., ITK-SNAP: https://www.itksnap.org/pmwiki/pmwiki.php, Siu Sai Look DICOM Viewer: https://xiaosaiviewer.com/) to load data for visual inspection of image quality and structure. Also, for users who need to program data processing and computational analyses, we recommend using pydicom ( https://pydicom.github.io/) or nibabel ( https://nipy.org/packages/nibabel/index.html) in a Python environment to read and Processing 3D image volumes; if needed, we can also provide sample visualisation source code to help users get started and implement customised image presentations. We also provide the code to handle the data processing at https://github.com/SuperD0122/LymphAtlas-.

To ensure the consistency of data preprocessing and segmentation analysis, users are advised to strictly follow the accompanying preprocessing instructions and record the software version and key parameters used to ensure the reproducibility of the experimental results. If you encounter problems during use or require customised code support, please refer to the accompanying source code examples or contact the dataset maintenance team for further assistance.

## 3.2 Continuous updating of data description

We currently collect a limited number of multimodal segmentation datasets to adequately reflect the diversity and complexity of lymphomas in clinical imaging. For this reason, we plan to establish a continuously updated data platform to supplement and expand the datasets on a regular basis to ensure that the data resources can keep pace with the actual clinical situation. Future work will screen and process the additional data through strict quality control processes and standardised annotation systems to ensure data consistency and high quality, thus providing more reliable support for subsequent model development and clinical applications.

In addition, we will disclose data updates on appropriate platforms and publish detailed update logs so that peer researchers can access the latest data in a timely manner and jointly advance the development of lymphoma image segmentation technology. Through this continuous updating mechanism, we expect to continuously enrich the diversity of the dataset and improve the generalisation ability of the algorithm, which will in turn provide a solid data foundation for accurate diagnosis and personalised treatment.

# 4. Summary

In this paper, a high-quality PET/CT multimodal lymphoma segmentation dataset was constructed and validated. A total of 483 FDG-PET/CT images of 220 lymphoma patients were retrospectively collected from March 2011 to May 2024, and were constructed into a multimodal dataset based on the nnUNet format after rigorous data preprocessing, alignment and expert annotation. After rigorous data pre-processing, alignment and expert annotation, a multimodal dataset is built based on the nnUNet format, which has been demonstrated to be highly accurate and robust in lesion localisation, morphology and metabolic characterisation in automatic segmentation algorithms. In the future, with the further expansion of data volume and the continuous innovation of imaging technology, this dataset will provide a solid foundation for the research and development of intelligent automatic segmentation methods and accurate diagnostic models based on multimodal image information, and it is expected to achieve all-around and three-dimensional portrayal of lymphoma lesions through the introduction of more imaging modalities (e.g., MRI, ultrasound, etc.) and the integration of multidimensional clinical data such as pathology and molecules; In addition, the in-depth integration of multimodal data and clinical indicators can provide more refined guidance for early detection, staging, efficacy assessment, and the formulation of personalised treatment plans, and promote the clinical translation and practical application of deep learning and AI in the field of precision medicine, so as to better address the current and future serious challenges in tumour management and treatment.